\documentclass[%
 reprint, showkeys,
 amsmath,amssymb,
 aps,
floatfix,
]{revtex4-2}

\usepackage{graphicx}
\usepackage{dcolumn}
\usepackage{bm}
\usepackage{xcolor}
\usepackage{textgreek}
\usepackage{gensymb}

\begin{document}

\preprint{APS/123-QED}

\title{Superconductivity in amorphous and crystalline Re-Lu films}

\author{Serafim Teknowijoyo}
\affiliation{Advanced Physics Laboratory, Institute for Quantum Studies, Chapman University, Burtonsville, MD 20866, USA}

\author{Armen Gulian}
\email[Corresponding author: ]{gulian@chapman.edu}
\affiliation{Advanced Physics Laboratory, Institute for Quantum Studies, Chapman University, Burtonsville, MD 20866, USA}


\begin{abstract}

We report on magnetron deposition and superconducting
properties of a novel superconducting material: rhenium-lutetium films on
sapphire substrates. Different compositions of Re$_{x}$Lu binary 
are explored from $x\approx 3.8$ to close to pure Re stoichiometry. 
The highest critical temperature, up to T$%
_{c}\approx $ 6.95 K, is obtained for $x\approx 10.5$. Depending on the
deposition conditions, polycrystalline or amorphous films are obtainable,
both of which are interesting for practice. Crystalline structure of
polycrystalline phase is identified using grazing incidence X-ray
diffractometry as a non-centrosymmetric superconductor. Superconducting
properties were characterized both resistively and magnetically.
Demonstration of superconductivity in this material justifies the point of
view that Lu plays a role of group 3 transition metal in period 6 of the
Periodic table of elements. In analogy with Re$_{0.82}$Nb$_{0.18}$, Re$_{6}$Ti, Re$_{6}$%
Hf and Re$_{6}$Zr, one can expect that crystalline Re--Lu is also
breaking the time-reversal symmetry (this still waits confirmation).
Magnetoresistivity and AC/DC susceptibility measurements allowed us to
determine H$_{c1}$ and H$_{c2}$ of these films, as well as estimate
coherence length $\xi (0)$ and magnetic penetration depth $\lambda _{L}(0)$.
We also provide information on surface morphology of these films.

\end{abstract}

\maketitle


\section{Introduction}

Non-centrosymmetric superconductors (NCS) with broken time-reversal symmetry
(TRS) are of great interest to contemporary ``hot topics" in
superconductivity. Namely, these materials \cite{Tokura18,Wakatsuki18,Hoshino18} provide the
unique opportunity to design simple and scalable superconducting devices
with nonreciprocal current control, such as diodes, transistors,
quadristors, etc. \cite{Ando20,Ideue20,BaumgartnerNN,Wu22,Strambini22,Morimoto18,Chahid23,TeknowijoyoQuad}. Application of these materials
inherently breaking NCS and TRS may abandon the necessity of external
magnetic fields and sophisticated nano-patterning (such as reported in 
\cite{Lyu21}). In Re-based compounds, such as Re-Nb \cite{Shang18}, Re-Ti \cite{Singh18},
Re-Hf \cite{Singh16} and Re-Zr \cite{Womack21}, muon
spectroscopy revealed broken time-reversal symmetry,
complementing their non-centrosymmetric origin. This adds momentum to
substantial interest in liquid-helium temperature superconductors which can
be used in quantum information and computation technologies \cite{Hazard19,Grunhaupt19,Niepce19}. 
These thin-film superconductors can be easily deposited, and they are
resistant to oxidation, have low resistivity, and/or are compatible with
high magnetic fields \cite{Samkharadze16,Borisov20}.

Rhenium itself belongs to transition metals, and in bulk form at ambient
conditions, it superconducts below T$_{c}\approx 1.7$ K \cite{Alekseevskii76,Roberts76,Song09}.
In a thin film form $T_{c}$ is higher \cite{Alekseevskii76,Ulhaq82,Frieberthauser70,Pappas18,TeknowijoyoRe} and can
reach values as high as 6 K. Compounds of Re with other transition metals
allowed material scientists to achieve not only higher values of $T_{c}$,
but also, as was mentioned above to demonstrate broken TRS in addition to NCS.
From this point of view it is interesting to explore superconductivity of
Re--Lu substance, since there is a widespread opinion that the lanthanide
Lu is closer to transition metals than La itself \cite{Scerri12}. However, Re--Lu
material has not been explored in bulk form, or in thin films.

To close this gap, we report here on superconducting properties of amorphous
and polycrystalline Re-Lu films with critical temperature up to about 7 K. We
studied morphology of the films, magnetotransport and magnetic
susceptibility which allowed us to estimate basic features of
superconducting state, such as the critical fields, coherence length, and
London penetration depth.

\section{Experimental details}

The Re-Lu films were prepared via magnetron sputtering in our ATC series UHV
Hybrid deposition system (AJA International, Inc.) with a base pressure of $1
\times\ 10^{-8}$ Torr. The Re target (ACI Alloys, Inc., 99.99\% purity) was
accommodated inside of a 1.5"\ DC gun. The Lu target (Heeger Materials,
Inc., purity Lu/TREM 99.99\%) was placed inside of a 2" DC gun. The sapphire
substrate (AdValue Technology, thickness 650 \textmu m, C-cut) was cleaned
thoroughly with isopropyl alcohol before it was mounted on the holder. In
our chamber's configuration, the substrate holder is at the center of the
chamber facing upwards, while the (five) sputtering guns are located at the
top. The substrate is rotated in plane throughout the whole deposition
process to ensure a homogeneous deposition layer over the whole surface. Our
predeposition \textit{in-situ} cleaning of the substrate typically involves heating it up to
900\degree{C} for 10 min followed by a gentle bombardment of Ar$^{+}$
ions at 600\degree {C} for 5-10 min using the Kauffman source at 45
degrees to the substrate surface. Then the temperature was raised to 
900\degree C and kept at that value for 30 min. Afterwards, the temperature was
reduced to 600\degree C and simultaneous deposition took place for
10 min, at pressure 3 - 4 mTorr, with gun power 250-260 W and anode voltage 605-460
V for Re, and with gun power 90-45 W and anode voltage correspondingly
325-275 {V} for Lu. Keeping the Re gun power constant, and
varying the sputtering power of Lu from case to case allowed us to vary the
values of $x$ in composition Re$_{x}$Lu (see Table I). After the
deposition, the temperature was raised back to 900\degree C for
in-situ annealing for 30 min and then cooled down to ambient temperature.
All the heating/cooling protocols consistently used a 30\degree C/min ramp
rate. The substrate was oriented to face the ion gun squarely.

\begin{table*}[]
    \centering
    \caption{Re/Lu stoichiometry ratio, its $T_c$ and the deposition parameters.}
    \begin{tabular}{lcccll}
Re/Lu &  P$_{Lu}$(W) & $T_{substr}$(\degree C) & Pressure (mTorr) & $T_c$(K) & Crystallinity \\ 
3.8   & 90           & 600                     & 3                & 5.25  & amorphous \\
7.4   & 55           & 30                     & 4                & 6.1   & amorphous \\
7.5   & 70           & 600                     & 3                & 5.8   & amorphous \\
10.5  & 55           & 600                     & 3                & 6.95  & polycrystalline \\
11.5  & 55           & 600                     & 3                & 6.75  & polycrystalline \\
$>$99 & 45           & 600                     & 4                & 6.3   & polycrystalline \\
\end{tabular}%
\begin{tabular}{l}
\end{tabular}
    \label{tab1}
\end{table*}

\section{Results}

Our initial choice of $x$ for examining Re$_{x}$Lu composition was $x\sim 4$, 
in analogy with the
well-known NCS superconductor Re$_{0.82}$Nb$_{0.18}$, known for its breaking of TRS.
The composition with $x \approx 3.8$ indeed turned out to be an amorphous
superconductor with $T_{c}\approx $ 5.3 K, see Fig. 1\textbf{(a)}.

\begin{figure}
    \centering
    \includegraphics[width=\linewidth]{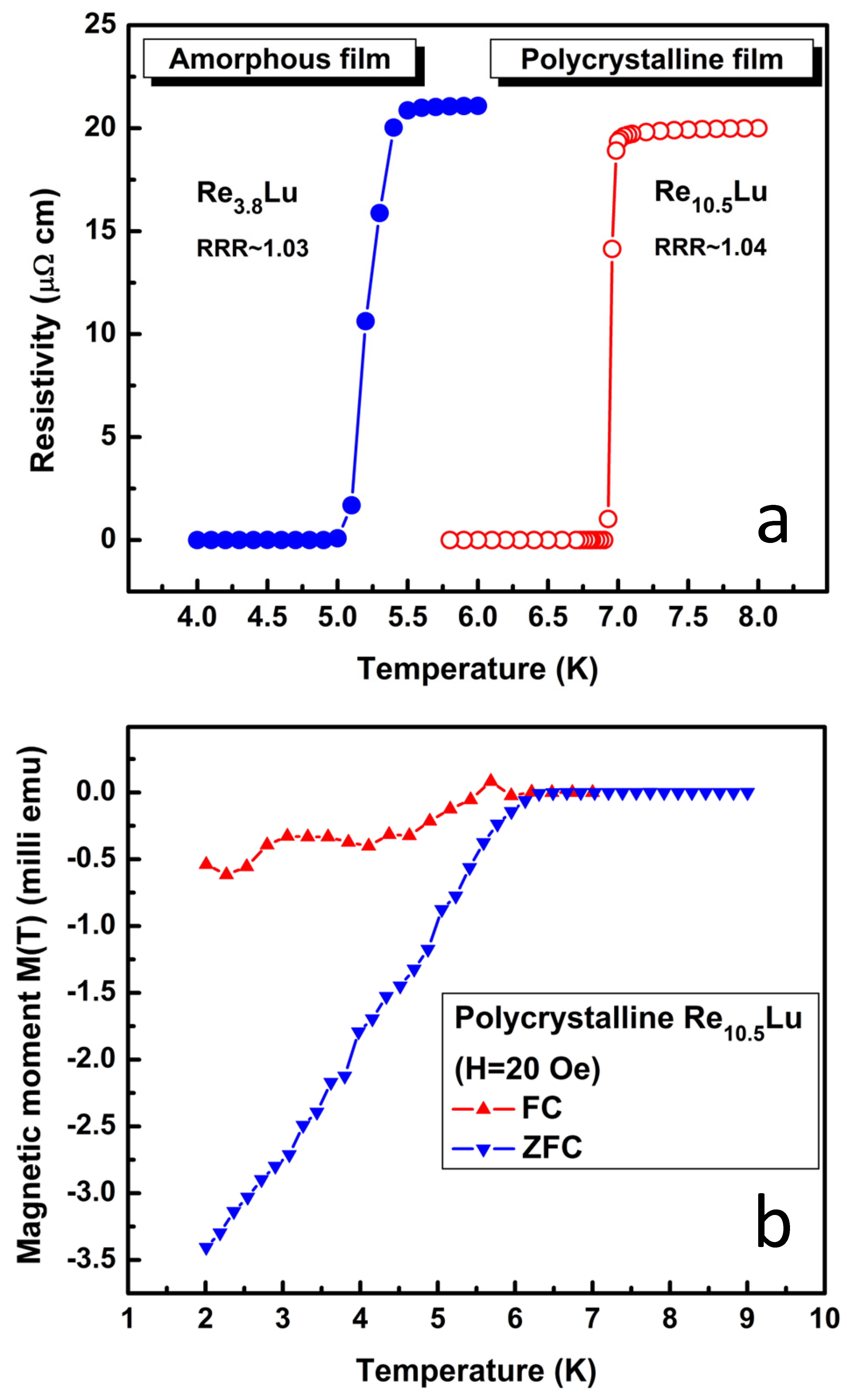}
    \caption{\textbf{(a)} Resistive superconducting transition in amorphous (at $x=3.8$) and
polycrystalline (at $x=10.5$) Re$_{x}$Lu films (Quantum Design PPMS); \textbf{(b)}
Meissner effect in polycrystalline Re$_{10.5}$Lu film.}
    \label{fig1}
\end{figure}

Lowering the relative concentration of the co-deposited Lu first increased
and then decreased the $T_{c}$, with the optimum $T_{c}\approx 7$ K
corresponding to $x\sim 10-11$ (shown also in Fig. 1). Though the normal state
resistivities of these two compositions are not much different, they have
very different surface morphologies, Fig. 2.

\begin{figure*}
    \centering
    \includegraphics[width=\linewidth]{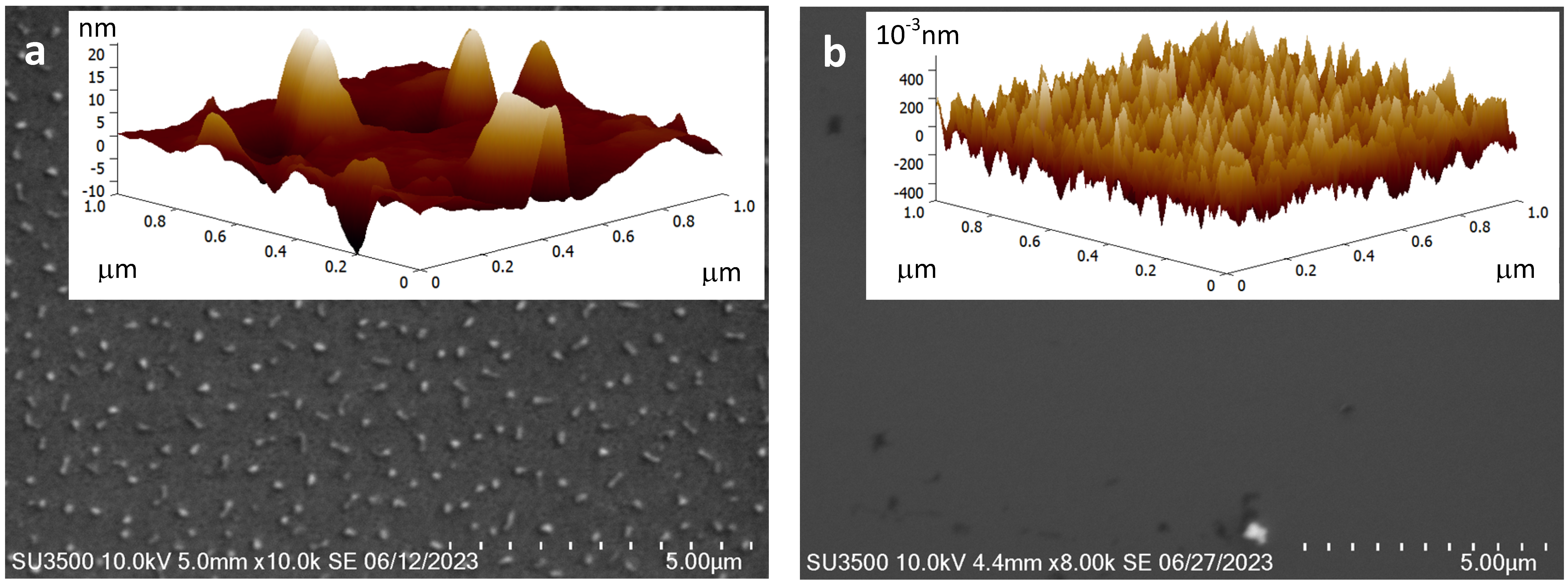}
    \caption{SEM (Hitachi SU3500) and AFM (insets, NT-MDT NTEGRA)
characterization of surface morphology of the films with $x=3.8$ \textbf{(a)} and $x=10.5$
\textbf{(b)}.}
    \label{fig2}
\end{figure*}
Comparative characteristics our films with various values of x are in Table I.    
To characterize the crystalline structure of our films we used grazing
incidence X-ray diffractometry which excludes the reflections of the
substrate. In this way it was recognized that the films with $x=3.8$ are
amorphous, and those with $x\approx 10$ are polycrystalline. 
In this report we will mainly focus on these two compositions. 
In the latter case,
using the diffractogram (Fig. 3) it is possible to determine the lattice
parameters of this novel substance.
\begin{figure}
    \centering
    \includegraphics[width=\linewidth]{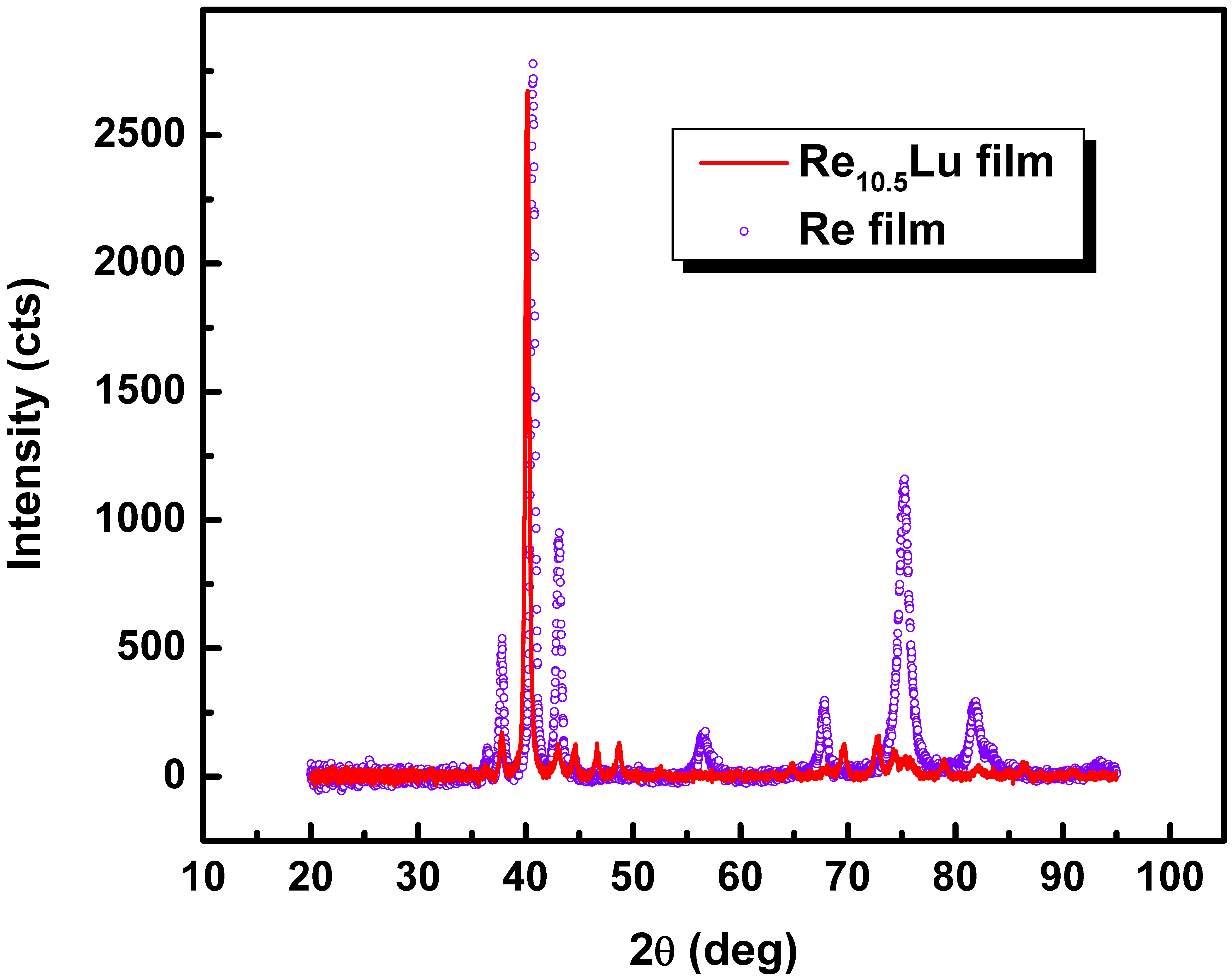}
    \caption{Grazing incidence X-ray diffraction pattern (Rigaku SmartLab) of
polycrystalline Re$_{10.5}$Lu film compared to that of pure Re film \cite{TeknowijoyoRe}.}
    \label{fig3}
\end{figure}
They are detailed in Table II. 
\begin{table*}[]
    \centering
    \caption{Crystal structure comparison}
    \begin{tabular}{llllll}
    Lattice parameters & \textit{a, \AA } & \textit{b, \AA } & \textit{c, \AA }
    & \textit{volume, \AA }$^{3}$ & \textit{space group} \\ 
    {Re}, bulk \cite{Alekseevskii76,Roberts76,Song09} & 2.761 & 2.761 & 4.458 & 29.430 & P63/mmc \\ 
    Re, film\cite{TeknowijoyoRe} & 2.782 & 2.782 & 4.484 & 30.053 & P63/mmc \\ 
    {Re}$_{10.5}${Lu} & 9.6004(11) & 9.6004(11) & 9.6004(11) & 
    884.86 & 217-43m%
    \end{tabular}
    \label{tab2}
\end{table*}
Magnetic and magnetotransport characterization of these films was
also performed (Quantum Design PPMS), Fig. 4 and Fig. 5.

\begin{figure*}
    \centering
    \includegraphics[width=\linewidth]{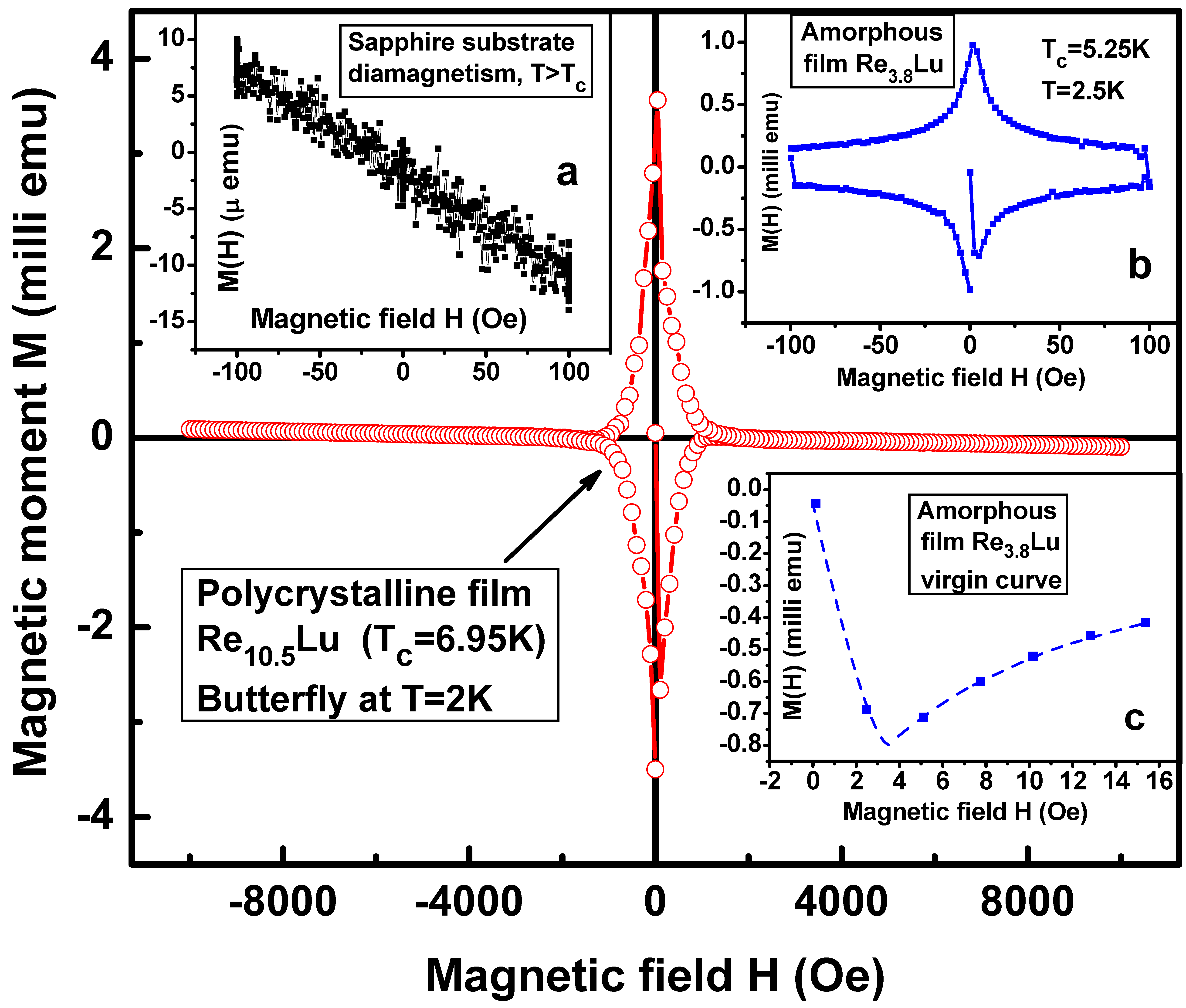}
    \caption{Characteristic ``butterfly"
pattern of polycrystalline film Re$_{10.5}$Lu. A noticeable clockwise rotation
of the butterfly is caused by the substrate diamagnetism (detailed in inset
\textbf{(a)}). Inset \textbf{(b)} demonstrates the ``butterfly" of Re$_{3.8}$Lu amorphous film.
Its virgin curve is shown in inset \textbf{c} (dashed lines are guides for eyes only).
}
    \label{fig4}
\end{figure*}

\begin{figure*}
    \centering
    \includegraphics[width=\linewidth]{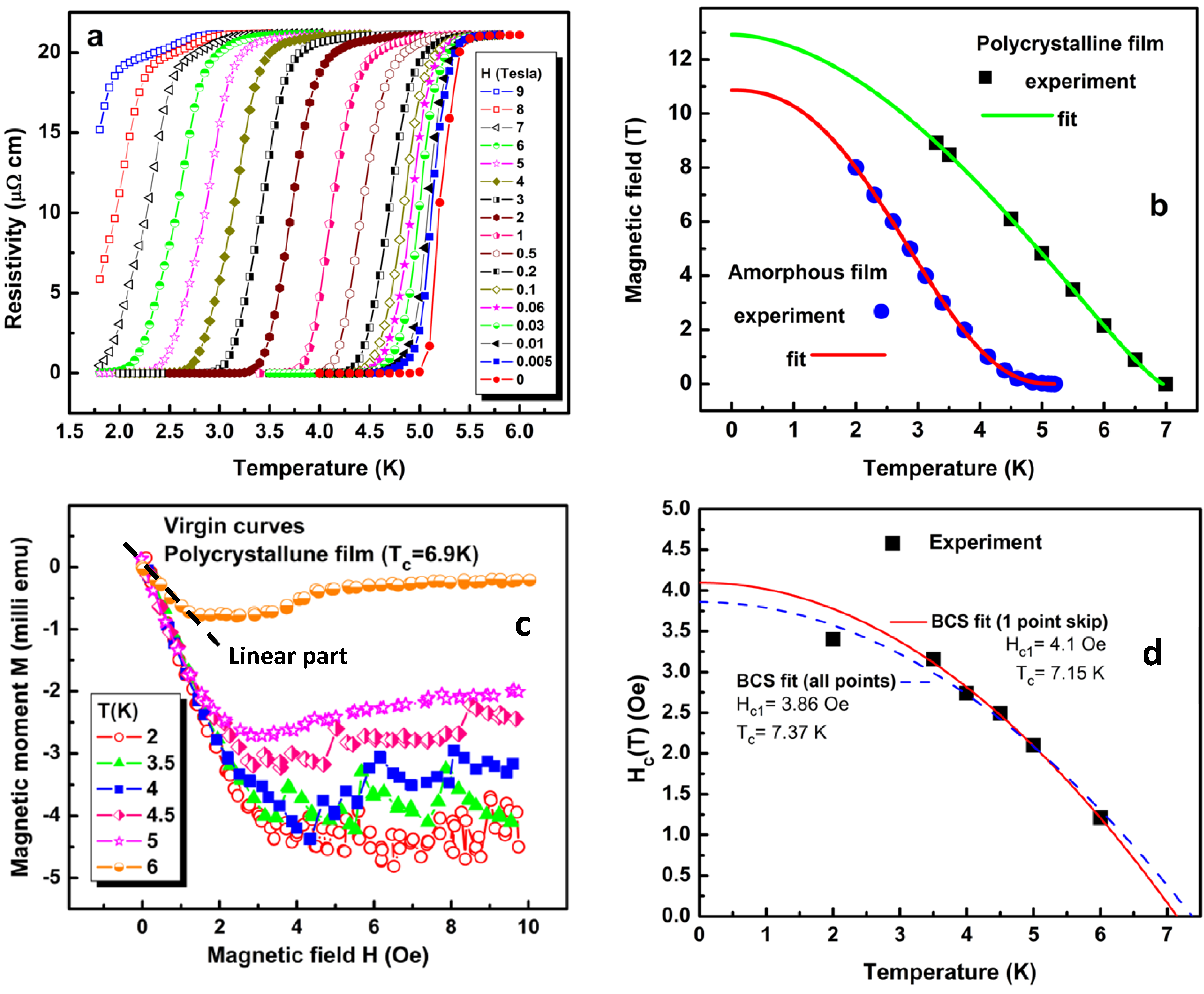}
    \caption{\textbf{(a)} Magnetotransport measurements for determining critical
field $H_{c2}$ vs. temperature in case of $x=3.8$ (amorphous) film. Similar
measurements were performed in case of $x=10.5$ (polycrystalline) film.  \textbf{(b)}
Determining $H_{c2}(T=0)$ for polycrystalline and amorphous films. \textbf{(c)}
Virgin curves of polycrystalline film at various temperatures for
determining the value of H$_{c1}$ (panel \textbf{d}). This value of $H_{c1}$ is
constructed in \textbf{(d)} using the linear part of experimental data (one such line
is shown in panel \textbf{c} - dashed line for $T=6$ K data) via modeling in accordance
with the relation $H_{c1}(T)=H_{c1}(0)[1-(T/T_{c})^{2}]$.}
    \label{fig5}
\end{figure*}

\section{Discussion}

As follows from Table I, both stoichiometric ratio and substrate's
temperature affect the crystalline properties of this material. Moreover,
the stoichiometric ratio itself depends on the substrate temperature. The
last entry in the table corresponds to less than 1\%(at.) of Lu in the
composition - we reached here the resolution limit of our energy-dispersive
spectrometer (Oxford Instruments X-Max$^{N}$). Meanwhile, as the special
investigation revealed \cite{TeknowijoyoRe}, pure Re films grown in the same conditions 
(600\degree C) are amorphous and do not superconduct down to 1.8 K, while being
grown on 30\degree C they do at 3.6 K. The role of the substrate-film interplay
is also important; for example, bulk Re never superconducts above 1.8 K
\cite{Alekseevskii76}.

Our samples' $H_{c2}$ (T) curves show different behavior compared to the
conventional BCS dependence $H_{c2} (T) = H_{c2}(0)[1 - (T/T_{c}
)^{2}]$. Therefore, the curves can instead be fitted using the expression 
$H_{c2} (T ) = H_{c2}(0)[1 - (T/T_{c} )^{p}]^{q}$ following 
\cite{Biswas11,Micnas90} where the exponents $p = q$ were chosen to be 
3/2. A slightly better fit to the data can be obtained when the
constraint on $p$ and $q$ are removed by choosing $p=1.8$,
$q=1.2$. This fit is shown in Fig. 5\textbf{(b)} which yields 
$H_{c2}\approx 13$ T. Using the Ginzburg-Landau relation $H_{c2} =\phi _{0}
/(2\pi \xi ^{2})$ , where $\phi _{0}= 2.068 \times 10^{-15}$ Wb is the
flux quantum \cite{Brandt88}, the estimated coherence length of our film with 
$T_{c}=6.95$ K is $\xi (0)= 5.05$ nm.

Combining $\xi (0)$ and the estimate for $H_{c1}$ from panel \textbf{(d)} in Fig. 5 (
$H_{c1}\sim 4$ Oe) \footnote{In this fitting procedure, the value of $T_{c}$ is a
free parameter. As follows from Fig. 5\textbf{(d)},
its value is close to the experimental $T_{c}\approx 7$ K} into the standard expression $H_{c1} = [\phi _{0}/(4\pi \lambda _{L}^{2})] [ln(\lambda _{L}/\xi ) +
0.12]$ \cite{TinkhamBook}, the estimate for our film's magnetic penetration depth is 
$\lambda _{L}(0) \approx 4.48$ nm. Finally, the Ginzburg-Landau parameter
can also be calculated, $\kappa =\lambda _{L}/\xi = 0.89$, which shows
that our film, being of type-II, is rather close to the theoretical boundary
separating type-I and type-II superconductors of $\kappa =1/ \sqrt{2}%
\approx 0.71$. This value of $\kappa $ for polycrystalline Re$_{10.5}$Lu
matches with that obtained for pure Re films \cite{TeknowijoyoRe}, though the
individual values of $\lambda _{L}$ and $\xi $ are different.

For the amorphous Re$_{3.8}$Lu with $T_{c}\approx 5.3$ K, the value of $H_{c2}$
is a bit smaller: $H_{c2}\approx 11$ T (second curve in Fig. 5\textbf{(b)}, obtained
with the fitting parameters $p=2.4$ and $q=2.9$) yielding an
estimate: $\xi =5.5$ nm. From Fig. 4\textbf{(c)}, $H_{c1}(T=2.5$K) $\approx 
$ 3 Oe. Taking into account that $H_{c1}(T=5.3$K) = 0, one can use
parabolic approximation used above and obtain an estimate $H_{c1}(0)%
\approx 3.88$ Oe \footnote{In this fitting procedure, the experimental value of $T_{c}$ is
used, and the parabola is enforced to go through through 2 points: 
$H_{c1} (T_{c}) = 0$ and $H_{c1}$(2.5K) $\approx $ 3 Oe. This method is less
accurate, however, it is satisfactory for estimates.}. 
This yields $\lambda _{L}(0) \approx
4.87$ nm and $\kappa = 0.89$, as in the case of polycrystalline films.

\section{Summary}

Thus, the idea that Lu can successfully play the role of a transition element
in Re--Lu compound is confirmed by this research. We obtained a new material, Re$_{x}$Lu 
($3.8 \leq x \leq 99$+). In particular, Re$_{10.5}$Lu exceeds the critical
temperature of known Re$_{6}$Hf, Re$_{6}$Zr and Re$_{6}$Ti. While these
superconductors have never been reported having $T_{c}>$ 6 K,
either in bulk or thin film form, Re$_{10.5}$Lu demonstrated $T_{c}\approx 7$ K.
By analogy, one can expect that this NCS material will also break
TRS. It will be very interesting to explore that property, though that goal
is beyond this paper. The indirect proof of broken TRS may be obtained by
effects related to nonreciprocal current control devices made of this
material: demonstration of nonreciprocity in absence of applied magnetic
field may serve as such a proof.

The simplicity of the described deposition method may facilitate the application of this material for wide
range of devices mentioned in Introduction. Also, the
information obtained by our research may provide grounds for further
fundamental studies based on band-structure computations of superconducting
state in Re-Lu materials to quantitatively explain the discovered features.
Finally, the parameters $\lambda_{L}$ and $\xi $ estimated above may be
used for modeling of phenomena in Re--Lu-based superconducting devices.

\begin{acknowledgments}

This research was supported by the ONR grants No.
N00014-21-1-2879 and No. N00014-20-1-2442.
We are grateful to Physics Art Frontiers for technical
assistance.

\end{acknowledgments}





\begin{thebibliography}{35}%
\makeatletter
\providecommand \@ifxundefined [1]{%
 \@ifx{#1\undefined}
}%
\providecommand \@ifnum [1]{%
 \ifnum #1\expandafter \@firstoftwo
 \else \expandafter \@secondoftwo
 \fi
}%
\providecommand \@ifx [1]{%
 \ifx #1\expandafter \@firstoftwo
 \else \expandafter \@secondoftwo
 \fi
}%
\providecommand \natexlab [1]{#1}%
\providecommand \enquote  [1]{``#1''}%
\providecommand \bibnamefont  [1]{#1}%
\providecommand \bibfnamefont [1]{#1}%
\providecommand \citenamefont [1]{#1}%
\providecommand \href@noop [0]{\@secondoftwo}%
\providecommand \href [0]{\begingroup \@sanitize@url \@href}%
\providecommand \@href[1]{\@@startlink{#1}\@@href}%
\providecommand \@@href[1]{\endgroup#1\@@endlink}%
\providecommand \@sanitize@url [0]{\catcode `\\12\catcode `\$12\catcode
  `\&12\catcode `\#12\catcode `\^12\catcode `\_12\catcode `\%12\relax}%
\providecommand \@@startlink[1]{}%
\providecommand \@@endlink[0]{}%
\providecommand \url  [0]{\begingroup\@sanitize@url \@url }%
\providecommand \@url [1]{\endgroup\@href {#1}{\urlprefix }}%
\providecommand \urlprefix  [0]{URL }%
\providecommand \Eprint [0]{\href }%
\providecommand \doibase [0]{https://doi.org/}%
\providecommand \selectlanguage [0]{\@gobble}%
\providecommand \bibinfo  [0]{\@secondoftwo}%
\providecommand \bibfield  [0]{\@secondoftwo}%
\providecommand \translation [1]{[#1]}%
\providecommand \BibitemOpen [0]{}%
\providecommand \bibitemStop [0]{}%
\providecommand \bibitemNoStop [0]{.\EOS\space}%
\providecommand \EOS [0]{\spacefactor3000\relax}%
\providecommand \BibitemShut  [1]{\csname bibitem#1\endcsname}%
\let\auto@bib@innerbib\@empty
\bibitem [{\citenamefont {Tokura}\ and\ \citenamefont
  {Nagaosa}(2018)}]{Tokura18}%
  \BibitemOpen
  \bibfield  {author} {\bibinfo {author} {\bibfnamefont {Y.}~\bibnamefont
  {Tokura}}\ and\ \bibinfo {author} {\bibfnamefont {N.}~\bibnamefont
  {Nagaosa}},\ }\bibfield  {title} {\bibinfo {title} {Nonreciprocal responses
  from non-centrosymmetric quantum materials},\ }\href@noop {} {\bibfield
  {journal} {\bibinfo  {journal} {Nature Communications}\ }\textbf {\bibinfo
  {volume} {9}},\ \bibinfo {pages} {3740} (\bibinfo {year} {2018})}\BibitemShut
  {NoStop}%
\bibitem [{\citenamefont {Wakatsuki}\ and\ \citenamefont
  {Nagaosa}(2018)}]{Wakatsuki18}%
  \BibitemOpen
  \bibfield  {author} {\bibinfo {author} {\bibfnamefont {R.}~\bibnamefont
  {Wakatsuki}}\ and\ \bibinfo {author} {\bibfnamefont {N.}~\bibnamefont
  {Nagaosa}},\ }\bibfield  {title} {\bibinfo {title} {{Nonreciprocal Current in
  Noncentrosymmetric Rashba Superconductors}},\ }\href@noop {} {\bibfield
  {journal} {\bibinfo  {journal} {Phys. Rev. Lett.}\ }\textbf {\bibinfo
  {volume} {121}},\ \bibinfo {pages} {026601} (\bibinfo {year}
  {2018})}\BibitemShut {NoStop}%
\bibitem [{\citenamefont {Hoshino}\ \emph {et~al.}(2018)\citenamefont
  {Hoshino}, \citenamefont {Wakatsuki}, \citenamefont {Hamamoto},\ and\
  \citenamefont {Nagaosa}}]{Hoshino18}%
  \BibitemOpen
  \bibfield  {author} {\bibinfo {author} {\bibfnamefont {S.}~\bibnamefont
  {Hoshino}}, \bibinfo {author} {\bibfnamefont {R.}~\bibnamefont {Wakatsuki}},
  \bibinfo {author} {\bibfnamefont {K.}~\bibnamefont {Hamamoto}},\ and\
  \bibinfo {author} {\bibfnamefont {N.}~\bibnamefont {Nagaosa}},\ }\bibfield
  {title} {\bibinfo {title} {Nonreciprocal charge transport in two-dimensional
  noncentrosymmetric superconductors},\ }\href@noop {} {\bibfield  {journal}
  {\bibinfo  {journal} {Phys. Rev. B}\ }\textbf {\bibinfo {volume} {98}},\
  \bibinfo {pages} {054510} (\bibinfo {year} {2018})}\BibitemShut {NoStop}%
\bibitem [{\citenamefont {Ando}\ \emph {et~al.}(2020)\citenamefont {Ando},
  \citenamefont {Miyasaka}, \citenamefont {Li}, \citenamefont {Ishizuka},
  \citenamefont {Arakawa}, \citenamefont {Shiota}, \citenamefont {Moriyama},
  \citenamefont {Yanase},\ and\ \citenamefont {Ono}}]{Ando20}%
  \BibitemOpen
  \bibfield  {author} {\bibinfo {author} {\bibfnamefont {F.}~\bibnamefont
  {Ando}}, \bibinfo {author} {\bibfnamefont {Y.}~\bibnamefont {Miyasaka}},
  \bibinfo {author} {\bibfnamefont {T.}~\bibnamefont {Li}}, \bibinfo {author}
  {\bibfnamefont {J.}~\bibnamefont {Ishizuka}}, \bibinfo {author}
  {\bibfnamefont {T.}~\bibnamefont {Arakawa}}, \bibinfo {author} {\bibfnamefont
  {Y.}~\bibnamefont {Shiota}}, \bibinfo {author} {\bibfnamefont
  {T.}~\bibnamefont {Moriyama}}, \bibinfo {author} {\bibfnamefont
  {Y.}~\bibnamefont {Yanase}},\ and\ \bibinfo {author} {\bibfnamefont
  {T.}~\bibnamefont {Ono}},\ }\bibfield  {title} {\bibinfo {title} {Observation
  of superconducting diode effect},\ }\href@noop {} {\bibfield  {journal}
  {\bibinfo  {journal} {Nature}\ }\textbf {\bibinfo {volume} {584}},\ \bibinfo
  {pages} {373} (\bibinfo {year} {2020})}\BibitemShut {NoStop}%
\bibitem [{\citenamefont {Ideue}\ and\ \citenamefont {Iwasa}(2020)}]{Ideue20}%
  \BibitemOpen
  \bibfield  {author} {\bibinfo {author} {\bibfnamefont {T.}~\bibnamefont
  {Ideue}}\ and\ \bibinfo {author} {\bibfnamefont {Y.}~\bibnamefont {Iwasa}},\
  }\bibfield  {title} {\bibinfo {title} {One-way supercurrent achieved in an
  electrically polar film},\ }\href@noop {} {\bibfield  {journal} {\bibinfo
  {journal} {Nature}\ }\textbf {\bibinfo {volume} {584}},\ \bibinfo {pages}
  {349} (\bibinfo {year} {2020})}\BibitemShut {NoStop}%
\bibitem [{\citenamefont {Baumgartner}\ \emph {et~al.}(2022)\citenamefont
  {Baumgartner}, \citenamefont {Fuchs}, \citenamefont {Costa}, \citenamefont
  {Reinhardt}, \citenamefont {Gronin}, \citenamefont {Gardner}, \citenamefont
  {Lindemann}, \citenamefont {Manfra}, \citenamefont {Faria~Junior},
  \citenamefont {Kochan}, \citenamefont {Fabian}, \citenamefont {Paradiso},\
  and\ \citenamefont {Strunk}}]{BaumgartnerNN}%
  \BibitemOpen
  \bibfield  {author} {\bibinfo {author} {\bibfnamefont {C.}~\bibnamefont
  {Baumgartner}}, \bibinfo {author} {\bibfnamefont {L.}~\bibnamefont {Fuchs}},
  \bibinfo {author} {\bibfnamefont {A.}~\bibnamefont {Costa}}, \bibinfo
  {author} {\bibfnamefont {S.}~\bibnamefont {Reinhardt}}, \bibinfo {author}
  {\bibfnamefont {S.}~\bibnamefont {Gronin}}, \bibinfo {author} {\bibfnamefont
  {G.~C.}\ \bibnamefont {Gardner}}, \bibinfo {author} {\bibfnamefont
  {T.}~\bibnamefont {Lindemann}}, \bibinfo {author} {\bibfnamefont {M.~J.}\
  \bibnamefont {Manfra}}, \bibinfo {author} {\bibfnamefont {P.~E.}\
  \bibnamefont {Faria~Junior}}, \bibinfo {author} {\bibfnamefont
  {D.}~\bibnamefont {Kochan}}, \bibinfo {author} {\bibfnamefont
  {J.}~\bibnamefont {Fabian}}, \bibinfo {author} {\bibfnamefont
  {N.}~\bibnamefont {Paradiso}},\ and\ \bibinfo {author} {\bibfnamefont
  {C.}~\bibnamefont {Strunk}},\ }\bibfield  {title} {\bibinfo {title}
  {{Supercurrent rectification and magnetochiral effects in symmetric Josephson
  junctions}},\ }\href@noop {} {\bibfield  {journal} {\bibinfo  {journal}
  {Nature Nanotechnology}\ }\textbf {\bibinfo {volume} {17}},\ \bibinfo {pages}
  {39} (\bibinfo {year} {2022})}\BibitemShut {NoStop}%
\bibitem [{\citenamefont {Wu}\ \emph {et~al.}(2022)\citenamefont {Wu},
  \citenamefont {Wang}, \citenamefont {Xu}, \citenamefont {Sivakumar},
  \citenamefont {Pasco}, \citenamefont {Filippozzi}, \citenamefont {Parkin},
  \citenamefont {Zeng}, \citenamefont {McQueen},\ and\ \citenamefont
  {Ali}}]{Wu22}%
  \BibitemOpen
  \bibfield  {author} {\bibinfo {author} {\bibfnamefont {H.}~\bibnamefont
  {Wu}}, \bibinfo {author} {\bibfnamefont {Y.}~\bibnamefont {Wang}}, \bibinfo
  {author} {\bibfnamefont {Y.}~\bibnamefont {Xu}}, \bibinfo {author}
  {\bibfnamefont {P.~K.}\ \bibnamefont {Sivakumar}}, \bibinfo {author}
  {\bibfnamefont {C.}~\bibnamefont {Pasco}}, \bibinfo {author} {\bibfnamefont
  {U.}~\bibnamefont {Filippozzi}}, \bibinfo {author} {\bibfnamefont {S.~S.~P.}\
  \bibnamefont {Parkin}}, \bibinfo {author} {\bibfnamefont {Y.-J.}\
  \bibnamefont {Zeng}}, \bibinfo {author} {\bibfnamefont {T.}~\bibnamefont
  {McQueen}},\ and\ \bibinfo {author} {\bibfnamefont {M.~N.}\ \bibnamefont
  {Ali}},\ }\bibfield  {title} {\bibinfo {title} {{The field-free Josephson
  diode in a van der Waals heterostructure}},\ }\href@noop {} {\bibfield
  {journal} {\bibinfo  {journal} {Nature}\ }\textbf {\bibinfo {volume} {604}},\
  \bibinfo {pages} {653} (\bibinfo {year} {2022})}\BibitemShut {NoStop}%
\bibitem [{\citenamefont {Strambini}\ \emph {et~al.}(2022)\citenamefont
  {Strambini}, \citenamefont {Spies}, \citenamefont {Ligato}, \citenamefont
  {Ili{\'{c}}}, \citenamefont {Rouco}, \citenamefont {Gonz{\'a}lez-Orellana},
  \citenamefont {Ilyn}, \citenamefont {Rogero}, \citenamefont {Bergeret},
  \citenamefont {Moodera}, \citenamefont {Virtanen}, \citenamefont
  {Heikkil{\"a}},\ and\ \citenamefont {Giazotto}}]{Strambini22}%
  \BibitemOpen
  \bibfield  {author} {\bibinfo {author} {\bibfnamefont {E.}~\bibnamefont
  {Strambini}}, \bibinfo {author} {\bibfnamefont {M.}~\bibnamefont {Spies}},
  \bibinfo {author} {\bibfnamefont {N.}~\bibnamefont {Ligato}}, \bibinfo
  {author} {\bibfnamefont {S.}~\bibnamefont {Ili{\'{c}}}}, \bibinfo {author}
  {\bibfnamefont {M.}~\bibnamefont {Rouco}}, \bibinfo {author} {\bibfnamefont
  {C.}~\bibnamefont {Gonz{\'a}lez-Orellana}}, \bibinfo {author} {\bibfnamefont
  {M.}~\bibnamefont {Ilyn}}, \bibinfo {author} {\bibfnamefont {C.}~\bibnamefont
  {Rogero}}, \bibinfo {author} {\bibfnamefont {F.~S.}\ \bibnamefont
  {Bergeret}}, \bibinfo {author} {\bibfnamefont {J.~S.}\ \bibnamefont
  {Moodera}}, \bibinfo {author} {\bibfnamefont {P.}~\bibnamefont {Virtanen}},
  \bibinfo {author} {\bibfnamefont {T.~T.}\ \bibnamefont {Heikkil{\"a}}},\ and\
  \bibinfo {author} {\bibfnamefont {F.}~\bibnamefont {Giazotto}},\ }\bibfield
  {title} {\bibinfo {title} {Superconducting spintronic tunnel diode},\
  }\href@noop {} {\bibfield  {journal} {\bibinfo  {journal} {Nature
  Communications}\ }\textbf {\bibinfo {volume} {13}},\ \bibinfo {pages} {2431}
  (\bibinfo {year} {2022})}\BibitemShut {NoStop}%
\bibitem [{\citenamefont {Morimoto}\ and\ \citenamefont
  {Nagaosa}(2018)}]{Morimoto18}%
  \BibitemOpen
  \bibfield  {author} {\bibinfo {author} {\bibfnamefont {T.}~\bibnamefont
  {Morimoto}}\ and\ \bibinfo {author} {\bibfnamefont {N.}~\bibnamefont
  {Nagaosa}},\ }\bibfield  {title} {\bibinfo {title} {Nonreciprocal current
  from electron interactions in noncentrosymmetric crystals: roles of time
  reversal symmetry and dissipation},\ }\href@noop {} {\bibfield  {journal}
  {\bibinfo  {journal} {Scientific Reports}\ }\textbf {\bibinfo {volume} {8}},\
  \bibinfo {pages} {2973} (\bibinfo {year} {2018})}\BibitemShut {NoStop}%
\bibitem [{\citenamefont {Chahid}\ \emph {et~al.}(2023)\citenamefont {Chahid},
  \citenamefont {Teknowijoyo}, \citenamefont {Mowgood},\ and\ \citenamefont
  {Gulian}}]{Chahid23}%
  \BibitemOpen
  \bibfield  {author} {\bibinfo {author} {\bibfnamefont {S.}~\bibnamefont
  {Chahid}}, \bibinfo {author} {\bibfnamefont {S.}~\bibnamefont {Teknowijoyo}},
  \bibinfo {author} {\bibfnamefont {I.}~\bibnamefont {Mowgood}},\ and\ \bibinfo
  {author} {\bibfnamefont {A.}~\bibnamefont {Gulian}},\ }\bibfield  {title}
  {\bibinfo {title} {{High-frequency diode effect in superconducting
  ${\mathrm{Nb}}_{3}\mathrm{Sn}$ microbridges}},\ }\href
  {https://doi.org/10.1103/PhysRevB.107.054506} {\bibfield  {journal} {\bibinfo
   {journal} {Phys. Rev. B}\ }\textbf {\bibinfo {volume} {107}},\ \bibinfo
  {pages} {054506} (\bibinfo {year} {2023})}\BibitemShut {NoStop}%
\bibitem [{\citenamefont {Teknowijoyo}\ \emph {et~al.}(2023)\citenamefont
  {Teknowijoyo}, \citenamefont {Chahid},\ and\ \citenamefont
  {Gulian}}]{TeknowijoyoQuad}%
  \BibitemOpen
  \bibfield  {author} {\bibinfo {author} {\bibfnamefont {S.}~\bibnamefont
  {Teknowijoyo}}, \bibinfo {author} {\bibfnamefont {S.}~\bibnamefont
  {Chahid}},\ and\ \bibinfo {author} {\bibfnamefont {A.}~\bibnamefont
  {Gulian}},\ }\bibfield  {title} {\bibinfo {title} {{Flux-Quanta Injection for
  Nonreciprocal Current Control in a Two-Dimensional Noncentrosymmetric
  Superconducting Structure}},\ }\href
  {https://doi.org/10.1103/PhysRevApplied.20.014055} {\bibfield  {journal}
  {\bibinfo  {journal} {Phys. Rev. Appl.}\ }\textbf {\bibinfo {volume} {20}},\
  \bibinfo {pages} {014055} (\bibinfo {year} {2023})}\BibitemShut {NoStop}%
\bibitem [{\citenamefont {Lyu}\ \emph {et~al.}(2021)\citenamefont {Lyu},
  \citenamefont {Jiang}, \citenamefont {Wang}, \citenamefont {Xiao},
  \citenamefont {Dong}, \citenamefont {Chen}, \citenamefont
  {Milo{\v{s}}evi{\'{c}}}, \citenamefont {Wang}, \citenamefont {Divan},
  \citenamefont {Pearson}, \citenamefont {Wu}, \citenamefont {Peeters},\ and\
  \citenamefont {Kwok}}]{Lyu21}%
  \BibitemOpen
  \bibfield  {author} {\bibinfo {author} {\bibfnamefont {Y.-Y.}\ \bibnamefont
  {Lyu}}, \bibinfo {author} {\bibfnamefont {J.}~\bibnamefont {Jiang}}, \bibinfo
  {author} {\bibfnamefont {Y.-L.}\ \bibnamefont {Wang}}, \bibinfo {author}
  {\bibfnamefont {Z.-L.}\ \bibnamefont {Xiao}}, \bibinfo {author}
  {\bibfnamefont {S.}~\bibnamefont {Dong}}, \bibinfo {author} {\bibfnamefont
  {Q.-H.}\ \bibnamefont {Chen}}, \bibinfo {author} {\bibfnamefont {M.~V.}\
  \bibnamefont {Milo{\v{s}}evi{\'{c}}}}, \bibinfo {author} {\bibfnamefont
  {H.}~\bibnamefont {Wang}}, \bibinfo {author} {\bibfnamefont {R.}~\bibnamefont
  {Divan}}, \bibinfo {author} {\bibfnamefont {J.~E.}\ \bibnamefont {Pearson}},
  \bibinfo {author} {\bibfnamefont {P.}~\bibnamefont {Wu}}, \bibinfo {author}
  {\bibfnamefont {F.~M.}\ \bibnamefont {Peeters}},\ and\ \bibinfo {author}
  {\bibfnamefont {W.-K.}\ \bibnamefont {Kwok}},\ }\bibfield  {title} {\bibinfo
  {title} {Superconducting diode effect via conformal-mapped nanoholes},\
  }\href@noop {} {\bibfield  {journal} {\bibinfo  {journal} {Nature
  Communications}\ }\textbf {\bibinfo {volume} {12}},\ \bibinfo {pages} {2703}
  (\bibinfo {year} {2021})}\BibitemShut {NoStop}%
\bibitem [{\citenamefont {Shang}\ \emph {et~al.}(2018)\citenamefont {Shang},
  \citenamefont {Smidman}, \citenamefont {Ghosh}, \citenamefont {Baines},
  \citenamefont {Chang}, \citenamefont {Gawryluk}, \citenamefont {Barker},
  \citenamefont {Singh}, \citenamefont {Paul}, \citenamefont {Balakrishnan},
  \citenamefont {Pomjakushina}, \citenamefont {Shi}, \citenamefont {Medarde},
  \citenamefont {Hillier}, \citenamefont {Yuan}, \citenamefont {Quintanilla},
  \citenamefont {Mesot},\ and\ \citenamefont {Shiroka}}]{Shang18}%
  \BibitemOpen
  \bibfield  {author} {\bibinfo {author} {\bibfnamefont {T.}~\bibnamefont
  {Shang}}, \bibinfo {author} {\bibfnamefont {M.}~\bibnamefont {Smidman}},
  \bibinfo {author} {\bibfnamefont {S.~K.}\ \bibnamefont {Ghosh}}, \bibinfo
  {author} {\bibfnamefont {C.}~\bibnamefont {Baines}}, \bibinfo {author}
  {\bibfnamefont {L.~J.}\ \bibnamefont {Chang}}, \bibinfo {author}
  {\bibfnamefont {D.~J.}\ \bibnamefont {Gawryluk}}, \bibinfo {author}
  {\bibfnamefont {J.~A.~T.}\ \bibnamefont {Barker}}, \bibinfo {author}
  {\bibfnamefont {R.~P.}\ \bibnamefont {Singh}}, \bibinfo {author}
  {\bibfnamefont {D.~M.}\ \bibnamefont {Paul}}, \bibinfo {author}
  {\bibfnamefont {G.}~\bibnamefont {Balakrishnan}}, \bibinfo {author}
  {\bibfnamefont {E.}~\bibnamefont {Pomjakushina}}, \bibinfo {author}
  {\bibfnamefont {M.}~\bibnamefont {Shi}}, \bibinfo {author} {\bibfnamefont
  {M.}~\bibnamefont {Medarde}}, \bibinfo {author} {\bibfnamefont {A.~D.}\
  \bibnamefont {Hillier}}, \bibinfo {author} {\bibfnamefont {H.~Q.}\
  \bibnamefont {Yuan}}, \bibinfo {author} {\bibfnamefont {J.}~\bibnamefont
  {Quintanilla}}, \bibinfo {author} {\bibfnamefont {J.}~\bibnamefont {Mesot}},\
  and\ \bibinfo {author} {\bibfnamefont {T.}~\bibnamefont {Shiroka}},\
  }\bibfield  {title} {\bibinfo {title} {{Time-Reversal Symmetry Breaking in
  Re-Based Superconductors}},\ }\href
  {https://doi.org/10.1103/PhysRevLett.121.257002} {\bibfield  {journal}
  {\bibinfo  {journal} {Phys. Rev. Lett.}\ }\textbf {\bibinfo {volume} {121}},\
  \bibinfo {pages} {257002} (\bibinfo {year} {2018})}\BibitemShut {NoStop}%
\bibitem [{\citenamefont {Singh}\ \emph {et~al.}(2018)\citenamefont {Singh},
  \citenamefont {K.~P.}, \citenamefont {Barker}, \citenamefont {Paul},
  \citenamefont {Hillier},\ and\ \citenamefont {Singh}}]{Singh18}%
  \BibitemOpen
  \bibfield  {author} {\bibinfo {author} {\bibfnamefont {D.}~\bibnamefont
  {Singh}}, \bibinfo {author} {\bibfnamefont {S.}~\bibnamefont {K.~P.}},
  \bibinfo {author} {\bibfnamefont {J.~A.~T.}\ \bibnamefont {Barker}}, \bibinfo
  {author} {\bibfnamefont {D.~M.}\ \bibnamefont {Paul}}, \bibinfo {author}
  {\bibfnamefont {A.~D.}\ \bibnamefont {Hillier}},\ and\ \bibinfo {author}
  {\bibfnamefont {R.~P.}\ \bibnamefont {Singh}},\ }\bibfield  {title} {\bibinfo
  {title} {{Time-reversal symmetry breaking in the noncentrosymmetric
  superconductor ${\mathrm{Re}}_{6}\mathrm{Ti}$}},\ }\href
  {https://doi.org/10.1103/PhysRevB.97.100505} {\bibfield  {journal} {\bibinfo
  {journal} {Phys. Rev. B}\ }\textbf {\bibinfo {volume} {97}},\ \bibinfo
  {pages} {100505} (\bibinfo {year} {2018})}\BibitemShut {NoStop}%
\bibitem [{\citenamefont {Singh}\ \emph {et~al.}(2016)\citenamefont {Singh},
  \citenamefont {Hillier}, \citenamefont {Thamizhavel},\ and\ \citenamefont
  {Singh}}]{Singh16}%
  \BibitemOpen
  \bibfield  {author} {\bibinfo {author} {\bibfnamefont {D.}~\bibnamefont
  {Singh}}, \bibinfo {author} {\bibfnamefont {A.~D.}\ \bibnamefont {Hillier}},
  \bibinfo {author} {\bibfnamefont {A.}~\bibnamefont {Thamizhavel}},\ and\
  \bibinfo {author} {\bibfnamefont {R.~P.}\ \bibnamefont {Singh}},\ }\bibfield
  {title} {\bibinfo {title} {{Superconducting properties of the
  noncentrosymmetric superconductor ${\mathrm{Re}}_{6}\mathrm{Hf}$}},\ }\href
  {https://doi.org/10.1103/PhysRevB.94.054515} {\bibfield  {journal} {\bibinfo
  {journal} {Phys. Rev. B}\ }\textbf {\bibinfo {volume} {94}},\ \bibinfo
  {pages} {054515} (\bibinfo {year} {2016})}\BibitemShut {NoStop}%
\bibitem [{\citenamefont {Womack}\ \emph {et~al.}(2021)\citenamefont {Womack},
  \citenamefont {Young}, \citenamefont {Browne}, \citenamefont {Catelani},
  \citenamefont {Jiang}, \citenamefont {Meletis},\ and\ \citenamefont
  {Adams}}]{Womack21}%
  \BibitemOpen
  \bibfield  {author} {\bibinfo {author} {\bibfnamefont {F.~N.}\ \bibnamefont
  {Womack}}, \bibinfo {author} {\bibfnamefont {D.~P.}\ \bibnamefont {Young}},
  \bibinfo {author} {\bibfnamefont {D.~A.}\ \bibnamefont {Browne}}, \bibinfo
  {author} {\bibfnamefont {G.}~\bibnamefont {Catelani}}, \bibinfo {author}
  {\bibfnamefont {J.}~\bibnamefont {Jiang}}, \bibinfo {author} {\bibfnamefont
  {E.~I.}\ \bibnamefont {Meletis}},\ and\ \bibinfo {author} {\bibfnamefont
  {P.~W.}\ \bibnamefont {Adams}},\ }\bibfield  {title} {\bibinfo {title}
  {{Extreme high-field superconductivity in thin Re films}},\ }\href
  {https://doi.org/10.1103/PhysRevB.103.024504} {\bibfield  {journal} {\bibinfo
   {journal} {Phys. Rev. B}\ }\textbf {\bibinfo {volume} {103}},\ \bibinfo
  {pages} {024504} (\bibinfo {year} {2021})}\BibitemShut {NoStop}%
\bibitem [{\citenamefont {Hazard}\ \emph {et~al.}(2019)\citenamefont {Hazard},
  \citenamefont {Gyenis}, \citenamefont {Di~Paolo}, \citenamefont {Asfaw},
  \citenamefont {Lyon}, \citenamefont {Blais},\ and\ \citenamefont
  {Houck}}]{Hazard19}%
  \BibitemOpen
  \bibfield  {author} {\bibinfo {author} {\bibfnamefont {T.~M.}\ \bibnamefont
  {Hazard}}, \bibinfo {author} {\bibfnamefont {A.}~\bibnamefont {Gyenis}},
  \bibinfo {author} {\bibfnamefont {A.}~\bibnamefont {Di~Paolo}}, \bibinfo
  {author} {\bibfnamefont {A.~T.}\ \bibnamefont {Asfaw}}, \bibinfo {author}
  {\bibfnamefont {S.~A.}\ \bibnamefont {Lyon}}, \bibinfo {author}
  {\bibfnamefont {A.}~\bibnamefont {Blais}},\ and\ \bibinfo {author}
  {\bibfnamefont {A.~A.}\ \bibnamefont {Houck}},\ }\bibfield  {title} {\bibinfo
  {title} {{Nanowire Superinductance Fluxonium Qubit}},\ }\href
  {https://doi.org/10.1103/PhysRevLett.122.010504} {\bibfield  {journal}
  {\bibinfo  {journal} {Phys. Rev. Lett.}\ }\textbf {\bibinfo {volume} {122}},\
  \bibinfo {pages} {010504} (\bibinfo {year} {2019})}\BibitemShut {NoStop}%
\bibitem [{\citenamefont {Gr\"unhaupt}\ \emph {et~al.}(2019)\citenamefont
  {Gr\"unhaupt}, \citenamefont {Spiecker}, \citenamefont {Gusenkova},
  \citenamefont {Maleeva}, \citenamefont {Skacel}, \citenamefont {Takmakov},
  \citenamefont {Valenti}, \citenamefont {Winkel}, \citenamefont {Rotzinger},
  \citenamefont {Wernsdorfer}, \citenamefont {Ustinov},\ and\ \citenamefont
  {Pop}}]{Grunhaupt19}%
  \BibitemOpen
  \bibfield  {author} {\bibinfo {author} {\bibfnamefont {L.}~\bibnamefont
  {Gr\"unhaupt}}, \bibinfo {author} {\bibfnamefont {M.}~\bibnamefont
  {Spiecker}}, \bibinfo {author} {\bibfnamefont {D.}~\bibnamefont {Gusenkova}},
  \bibinfo {author} {\bibfnamefont {N.}~\bibnamefont {Maleeva}}, \bibinfo
  {author} {\bibfnamefont {S.~T.}\ \bibnamefont {Skacel}}, \bibinfo {author}
  {\bibfnamefont {I.}~\bibnamefont {Takmakov}}, \bibinfo {author}
  {\bibfnamefont {F.}~\bibnamefont {Valenti}}, \bibinfo {author} {\bibfnamefont
  {P.}~\bibnamefont {Winkel}}, \bibinfo {author} {\bibfnamefont
  {H.}~\bibnamefont {Rotzinger}}, \bibinfo {author} {\bibfnamefont
  {W.}~\bibnamefont {Wernsdorfer}}, \bibinfo {author} {\bibfnamefont {A.~V.}\
  \bibnamefont {Ustinov}},\ and\ \bibinfo {author} {\bibfnamefont {I.~M.}\
  \bibnamefont {Pop}},\ }\bibfield  {title} {\bibinfo {title} {Granular
  aluminium as a superconducting material for high-impedance quantum
  circuits},\ }\href {https://doi.org/10.1038/s41563-019-0350-3} {\bibfield
  {journal} {\bibinfo  {journal} {Nature Materials}\ }\textbf {\bibinfo
  {volume} {18}},\ \bibinfo {pages} {816} (\bibinfo {year} {2019})}\BibitemShut
  {NoStop}%
\bibitem [{\citenamefont {Niepce}\ \emph {et~al.}(2019)\citenamefont {Niepce},
  \citenamefont {Burnett},\ and\ \citenamefont {Bylander}}]{Niepce19}%
  \BibitemOpen
  \bibfield  {author} {\bibinfo {author} {\bibfnamefont {D.}~\bibnamefont
  {Niepce}}, \bibinfo {author} {\bibfnamefont {J.}~\bibnamefont {Burnett}},\
  and\ \bibinfo {author} {\bibfnamefont {J.}~\bibnamefont {Bylander}},\
  }\bibfield  {title} {\bibinfo {title} {{High Kinetic Inductance
  $\mathrm{Nb}\mathrm{N}$ Nanowire Superinductors}},\ }\href
  {https://doi.org/10.1103/PhysRevApplied.11.044014} {\bibfield  {journal}
  {\bibinfo  {journal} {Phys. Rev. Appl.}\ }\textbf {\bibinfo {volume} {11}},\
  \bibinfo {pages} {044014} (\bibinfo {year} {2019})}\BibitemShut {NoStop}%
\bibitem [{\citenamefont {Samkharadze}\ \emph {et~al.}(2016)\citenamefont
  {Samkharadze}, \citenamefont {Bruno}, \citenamefont {Scarlino}, \citenamefont
  {Zheng}, \citenamefont {DiVincenzo}, \citenamefont {DiCarlo},\ and\
  \citenamefont {Vandersypen}}]{Samkharadze16}%
  \BibitemOpen
  \bibfield  {author} {\bibinfo {author} {\bibfnamefont {N.}~\bibnamefont
  {Samkharadze}}, \bibinfo {author} {\bibfnamefont {A.}~\bibnamefont {Bruno}},
  \bibinfo {author} {\bibfnamefont {P.}~\bibnamefont {Scarlino}}, \bibinfo
  {author} {\bibfnamefont {G.}~\bibnamefont {Zheng}}, \bibinfo {author}
  {\bibfnamefont {D.~P.}\ \bibnamefont {DiVincenzo}}, \bibinfo {author}
  {\bibfnamefont {L.}~\bibnamefont {DiCarlo}},\ and\ \bibinfo {author}
  {\bibfnamefont {L.~M.~K.}\ \bibnamefont {Vandersypen}},\ }\bibfield  {title}
  {\bibinfo {title} {{High-Kinetic-Inductance Superconducting Nanowire
  Resonators for Circuit QED in a Magnetic Field}},\ }\href
  {https://doi.org/10.1103/PhysRevApplied.5.044004} {\bibfield  {journal}
  {\bibinfo  {journal} {Phys. Rev. Appl.}\ }\textbf {\bibinfo {volume} {5}},\
  \bibinfo {pages} {044004} (\bibinfo {year} {2016})}\BibitemShut {NoStop}%
\bibitem [{\citenamefont {Borisov}\ \emph {et~al.}(2020)\citenamefont
  {Borisov}, \citenamefont {Rieger}, \citenamefont {Winkel}, \citenamefont
  {Henriques}, \citenamefont {Valenti}, \citenamefont {Ionita}, \citenamefont
  {Wessbecher}, \citenamefont {Spiecker}, \citenamefont {Gusenkova},
  \citenamefont {Pop},\ and\ \citenamefont {Wernsdorfer}}]{Borisov20}%
  \BibitemOpen
  \bibfield  {author} {\bibinfo {author} {\bibfnamefont {K.}~\bibnamefont
  {Borisov}}, \bibinfo {author} {\bibfnamefont {D.}~\bibnamefont {Rieger}},
  \bibinfo {author} {\bibfnamefont {P.}~\bibnamefont {Winkel}}, \bibinfo
  {author} {\bibfnamefont {F.}~\bibnamefont {Henriques}}, \bibinfo {author}
  {\bibfnamefont {F.}~\bibnamefont {Valenti}}, \bibinfo {author} {\bibfnamefont
  {A.}~\bibnamefont {Ionita}}, \bibinfo {author} {\bibfnamefont
  {M.}~\bibnamefont {Wessbecher}}, \bibinfo {author} {\bibfnamefont
  {M.}~\bibnamefont {Spiecker}}, \bibinfo {author} {\bibfnamefont
  {D.}~\bibnamefont {Gusenkova}}, \bibinfo {author} {\bibfnamefont {I.~M.}\
  \bibnamefont {Pop}},\ and\ \bibinfo {author} {\bibfnamefont {W.}~\bibnamefont
  {Wernsdorfer}},\ }\bibfield  {title} {\bibinfo {title} {{Superconducting
  granular aluminum resonators resilient to magnetic fields up to 1 Tesla}},\
  }\href {https://doi.org/10.1063/5.0018012} {\bibfield  {journal} {\bibinfo
  {journal} {Applied Physics Letters}\ }\textbf {\bibinfo {volume} {117}},\
  \bibinfo {pages} {120502} (\bibinfo {year} {2020})}\BibitemShut {NoStop}%
\bibitem [{\citenamefont {Alekseevskii}\ \emph {et~al.}(1967)\citenamefont
  {Alekseevskii}, \citenamefont {Mikheeva},\ and\ \citenamefont
  {Tulina}}]{Alekseevskii76}%
  \BibitemOpen
  \bibfield  {author} {\bibinfo {author} {\bibfnamefont {N.~E.}\ \bibnamefont
  {Alekseevskii}}, \bibinfo {author} {\bibfnamefont {M.~N.}\ \bibnamefont
  {Mikheeva}},\ and\ \bibinfo {author} {\bibfnamefont {N.~A.}\ \bibnamefont
  {Tulina}},\ }\bibfield  {title} {\bibinfo {title} {The superconducting
  properties of rhenium},\ }\href@noop {} {\bibfield  {journal} {\bibinfo
  {journal} {Sov. Phys. JETP}\ }\textbf {\bibinfo {volume} {25}},\ \bibinfo
  {pages} {575} (\bibinfo {year} {1967})},\ \bibinfo {note} {[Zh. Eksp. Teor.
  Fiz. 52, 875(1967)]}\BibitemShut {NoStop}%
\bibitem [{\citenamefont {Roberts}(1976)}]{Roberts76}%
  \BibitemOpen
  \bibfield  {author} {\bibinfo {author} {\bibfnamefont {B.~W.}\ \bibnamefont
  {Roberts}},\ }\bibfield  {title} {\bibinfo {title} {{Survey of
  superconductive materials and critical evaluation of selected properties}},\
  }\href {https://doi.org/10.1063/1.555540} {\bibfield  {journal} {\bibinfo
  {journal} {Journal of Physical and Chemical Reference Data}\ }\textbf
  {\bibinfo {volume} {5}},\ \bibinfo {pages} {581} (\bibinfo {year}
  {1976})}\BibitemShut {NoStop}%
\bibitem [{\citenamefont {Song}\ \emph {et~al.}(2009)\citenamefont {Song},
  \citenamefont {Heitmann}, \citenamefont {DeFeo}, \citenamefont {Yu},
  \citenamefont {McDermott}, \citenamefont {Neeley}, \citenamefont {Martinis},\
  and\ \citenamefont {Plourde}}]{Song09}%
  \BibitemOpen
  \bibfield  {author} {\bibinfo {author} {\bibfnamefont {C.}~\bibnamefont
  {Song}}, \bibinfo {author} {\bibfnamefont {T.~W.}\ \bibnamefont {Heitmann}},
  \bibinfo {author} {\bibfnamefont {M.~P.}\ \bibnamefont {DeFeo}}, \bibinfo
  {author} {\bibfnamefont {K.}~\bibnamefont {Yu}}, \bibinfo {author}
  {\bibfnamefont {R.}~\bibnamefont {McDermott}}, \bibinfo {author}
  {\bibfnamefont {M.}~\bibnamefont {Neeley}}, \bibinfo {author} {\bibfnamefont
  {J.~M.}\ \bibnamefont {Martinis}},\ and\ \bibinfo {author} {\bibfnamefont
  {B.~L.~T.}\ \bibnamefont {Plourde}},\ }\bibfield  {title} {\bibinfo {title}
  {{Microwave response of vortices in superconducting thin films of Re and
  Al}},\ }\href {https://doi.org/10.1103/PhysRevB.79.174512} {\bibfield
  {journal} {\bibinfo  {journal} {Phys. Rev. B}\ }\textbf {\bibinfo {volume}
  {79}},\ \bibinfo {pages} {174512} (\bibinfo {year} {2009})}\BibitemShut
  {NoStop}%
\bibitem [{\citenamefont {Haq}\ and\ \citenamefont {Meyer}(1982)}]{Ulhaq82}%
  \BibitemOpen
  \bibfield  {author} {\bibinfo {author} {\bibfnamefont {A.}~\bibnamefont
  {Haq}}\ and\ \bibinfo {author} {\bibfnamefont {O.}~\bibnamefont {Meyer}},\
  }\bibfield  {title} {\bibinfo {title} {Electrical and superconducting
  properties of rhenium thin films},\ }\href
  {https://doi.org/https://doi.org/10.1016/0040-6090(82)90504-1} {\bibfield
  {journal} {\bibinfo  {journal} {Thin Solid Films}\ }\textbf {\bibinfo
  {volume} {94}},\ \bibinfo {pages} {119} (\bibinfo {year} {1982})}\BibitemShut
  {NoStop}%
\bibitem [{\citenamefont {Frieberthauser}\ and\ \citenamefont
  {Notarys}(1970)}]{Frieberthauser70}%
  \BibitemOpen
  \bibfield  {author} {\bibinfo {author} {\bibfnamefont {P.~E.}\ \bibnamefont
  {Frieberthauser}}\ and\ \bibinfo {author} {\bibfnamefont {H.~A.}\
  \bibnamefont {Notarys}},\ }\bibfield  {title} {\bibinfo {title} {{Electrical
  Properties and Superconductivity of Rhenium and Molybdenum Films}},\ }\href
  {https://doi.org/10.1116/1.1315371} {\bibfield  {journal} {\bibinfo
  {journal} {Journal of Vacuum Science and Technology}\ }\textbf {\bibinfo
  {volume} {7}},\ \bibinfo {pages} {485} (\bibinfo {year} {1970})}\BibitemShut
  {NoStop}%
\bibitem [{\citenamefont {Pappas}\ \emph {et~al.}(2018)\citenamefont {Pappas},
  \citenamefont {David}, \citenamefont {Lake}, \citenamefont {Bal},
  \citenamefont {Goldfarb}, \citenamefont {Hite}, \citenamefont {Kim},
  \citenamefont {Ku}, \citenamefont {Long}, \citenamefont {McRae},
  \citenamefont {Pappas}, \citenamefont {Roshko}, \citenamefont {Wen},
  \citenamefont {Plourde}, \citenamefont {Arslan},\ and\ \citenamefont
  {Wu}}]{Pappas18}%
  \BibitemOpen
  \bibfield  {author} {\bibinfo {author} {\bibfnamefont {D.~P.}\ \bibnamefont
  {Pappas}}, \bibinfo {author} {\bibfnamefont {D.~E.}\ \bibnamefont {David}},
  \bibinfo {author} {\bibfnamefont {R.~E.}\ \bibnamefont {Lake}}, \bibinfo
  {author} {\bibfnamefont {M.}~\bibnamefont {Bal}}, \bibinfo {author}
  {\bibfnamefont {R.~B.}\ \bibnamefont {Goldfarb}}, \bibinfo {author}
  {\bibfnamefont {D.~A.}\ \bibnamefont {Hite}}, \bibinfo {author}
  {\bibfnamefont {E.}~\bibnamefont {Kim}}, \bibinfo {author} {\bibfnamefont
  {H.-S.}\ \bibnamefont {Ku}}, \bibinfo {author} {\bibfnamefont {J.~L.}\
  \bibnamefont {Long}}, \bibinfo {author} {\bibfnamefont {C.~R.~H.}\
  \bibnamefont {McRae}}, \bibinfo {author} {\bibfnamefont {L.~D.}\ \bibnamefont
  {Pappas}}, \bibinfo {author} {\bibfnamefont {A.}~\bibnamefont {Roshko}},
  \bibinfo {author} {\bibfnamefont {J.~G.}\ \bibnamefont {Wen}}, \bibinfo
  {author} {\bibfnamefont {B.~L.~T.}\ \bibnamefont {Plourde}}, \bibinfo
  {author} {\bibfnamefont {I.}~\bibnamefont {Arslan}},\ and\ \bibinfo {author}
  {\bibfnamefont {X.}~\bibnamefont {Wu}},\ }\bibfield  {title} {\bibinfo
  {title} {{Enhanced superconducting transition temperature in electroplated
  rhenium}},\ }\href {https://doi.org/10.1063/1.5027104} {\bibfield  {journal}
  {\bibinfo  {journal} {Applied Physics Letters}\ }\textbf {\bibinfo {volume}
  {112}},\ \bibinfo {pages} {182601} (\bibinfo {year} {2018})}\BibitemShut
  {NoStop}%
\bibitem [{\citenamefont {Teknowijoyo}\ and\ \citenamefont
  {Gulian}(2023)}]{TeknowijoyoRe}%
  \BibitemOpen
  \bibfield  {author} {\bibinfo {author} {\bibfnamefont {S.}~\bibnamefont
  {Teknowijoyo}}\ and\ \bibinfo {author} {\bibfnamefont {A.}~\bibnamefont
  {Gulian}},\ }\bibfield  {title} {\bibinfo {title} {{Superconducting
  polycrystalline rhenium films deposited at room temperature}},\ }\href@noop
  {} {\bibfield  {journal} {\bibinfo  {journal} {Opt. Mem. Neur. Networks}\
  }\textbf {\bibinfo {volume} {32}} (\bibinfo {year} {2023})}\BibitemShut
  {NoStop}%
\bibitem [{\citenamefont {Scerri}(2012)}]{Scerri12}%
  \BibitemOpen
  \bibfield  {author} {\bibinfo {author} {\bibfnamefont {E.}~\bibnamefont
  {Scerri}},\ }\bibfield  {title} {\bibinfo {title} {{Mendeleev's Periodic
  Table Is Finally Completed and What To Do about Group 3?}},\ }\href
  {https://doi.org/doi:10.1515/ci.2012.34.4.28} {\bibfield  {journal} {\bibinfo
   {journal} {Chemistry International -- Newsmagazine for IUPAC}\ }\textbf
  {\bibinfo {volume} {34}},\ \bibinfo {pages} {28} (\bibinfo {year}
  {2012})}\BibitemShut {NoStop}%
\bibitem [{\citenamefont {Biswas}\ \emph {et~al.}(2011)\citenamefont {Biswas},
  \citenamefont {Lees}, \citenamefont {Hillier}, \citenamefont {Smith},
  \citenamefont {Marshall},\ and\ \citenamefont {Paul}}]{Biswas11}%
  \BibitemOpen
  \bibfield  {author} {\bibinfo {author} {\bibfnamefont {P.~K.}\ \bibnamefont
  {Biswas}}, \bibinfo {author} {\bibfnamefont {M.~R.}\ \bibnamefont {Lees}},
  \bibinfo {author} {\bibfnamefont {A.~D.}\ \bibnamefont {Hillier}}, \bibinfo
  {author} {\bibfnamefont {R.~I.}\ \bibnamefont {Smith}}, \bibinfo {author}
  {\bibfnamefont {W.~G.}\ \bibnamefont {Marshall}},\ and\ \bibinfo {author}
  {\bibfnamefont {D.~M.}\ \bibnamefont {Paul}},\ }\bibfield  {title} {\bibinfo
  {title} {{Structure and superconductivity of two different phases of
  Re${}_{3}$W}},\ }\href {https://doi.org/10.1103/PhysRevB.84.184529}
  {\bibfield  {journal} {\bibinfo  {journal} {Phys. Rev. B}\ }\textbf {\bibinfo
  {volume} {84}},\ \bibinfo {pages} {184529} (\bibinfo {year}
  {2011})}\BibitemShut {NoStop}%
\bibitem [{\citenamefont {Micnas}\ \emph {et~al.}(1990)\citenamefont {Micnas},
  \citenamefont {Ranninger},\ and\ \citenamefont {Robaszkiewicz}}]{Micnas90}%
  \BibitemOpen
  \bibfield  {author} {\bibinfo {author} {\bibfnamefont {R.}~\bibnamefont
  {Micnas}}, \bibinfo {author} {\bibfnamefont {J.}~\bibnamefont {Ranninger}},\
  and\ \bibinfo {author} {\bibfnamefont {S.}~\bibnamefont {Robaszkiewicz}},\
  }\bibfield  {title} {\bibinfo {title} {Superconductivity in narrow-band
  systems with local nonretarded attractive interactions},\ }\href
  {https://doi.org/10.1103/RevModPhys.62.113} {\bibfield  {journal} {\bibinfo
  {journal} {Rev. Mod. Phys.}\ }\textbf {\bibinfo {volume} {62}},\ \bibinfo
  {pages} {113} (\bibinfo {year} {1990})}\BibitemShut {NoStop}%
\bibitem [{\citenamefont {Brandt}(1988)}]{Brandt88}%
  \BibitemOpen
  \bibfield  {author} {\bibinfo {author} {\bibfnamefont {E.~H.}\ \bibnamefont
  {Brandt}},\ }\bibfield  {title} {\bibinfo {title} {{Flux distribution and
  penetration depth measured by muon spin rotation in high-${T}_{c}$
  superconductors}},\ }\href {https://doi.org/10.1103/PhysRevB.37.2349}
  {\bibfield  {journal} {\bibinfo  {journal} {Phys. Rev. B}\ }\textbf {\bibinfo
  {volume} {37}},\ \bibinfo {pages} {2349} (\bibinfo {year}
  {1988})}\BibitemShut {NoStop}%
\bibitem [{Note1()}]{Note1}%
  \BibitemOpen
  \bibinfo {note} {In this fitting procedure, the value of $T_{c}$ is a free
  parameter. As follows from Fig. 5\protect \textbf {(d)}, its value is close
  to the experimental $T_{c}\approx 7$ K}\BibitemShut {NoStop}%
\bibitem [{\citenamefont {Tinkham}(1975)}]{TinkhamBook}%
  \BibitemOpen
  \bibfield  {author} {\bibinfo {author} {\bibfnamefont {M.}~\bibnamefont
  {Tinkham}},\ }\href@noop {} {\emph {\bibinfo {title} {{Introduction to
  Superconductivity}}}},\ International series in pure and applied physics\
  (\bibinfo  {publisher} {McGraw-Hill},\ \bibinfo {address} {New York},\
  \bibinfo {year} {1975})\BibitemShut {NoStop}%
\bibitem [{Note2()}]{Note2}%
  \BibitemOpen
  \bibinfo {note} {In this fitting procedure, the experimental value of $T_{c}$
  is used, and the parabola is enforced to go through through 2 points: $H_{c1}
  (T_{c}) = 0$ and $H_{c1}$(2.5K) $\approx $ 3 Oe. This method is less
  accurate, however, it is satisfactory for estimates.}\BibitemShut {Stop}%
\end{thebibliography}

\providecommand{\noopsort}[1]{}\providecommand{\singleletter}[1]{#1}%

\end{document}